\documentclass[aps,prb,twocolumn,,showpacs,superscriptaddress]{revtex4-1}

\usepackage{graphicx} 
\usepackage{amsmath}
\usepackage{bm}
\usepackage[colorlinks]{hyperref}
\usepackage{color}

\newcommand{\bj}{\bar{\jmath}}
\newcommand{\bdelta}{\bm \delta}

\newcommand{\EqLabel}[1]{\label{#1}} \newcommand{\mb}[1]{\mathbf{#1}}

\begin{document}
 
\title{Aharonov-Bohm interference for a hole in a two-dimensional Ising
  antiferromagnet in a transverse magnetic field}

\author{Mona Berciu}

\affiliation{Department of Physics and Astronomy, University of
  British Columbia, Vancouver, BC V6T 1Z1, Canada }
\affiliation{Quantum Matter Institute, University of
  British Columbia, Vancouver, BC V6T 1Z4, Canada }

\author{Holger Fehske}

\affiliation{Institut f\"ur Physik, Ernst-Moritz-Arndt-Universit\"at
  Greifswald, D-17487 Greifswald, Germany }

\date{\today}
 
\begin{abstract} 
We show that a proper consideration of the contribution of 
Trugman loops leads to a fairly low effective mass for a hole
moving in a square lattice Ising antiferromagnet, if the
bare hopping and the 
exchange energy scales are comparable. This contradicts the general
view that because of the absence of spin fluctuations, this
effective mass must be extremely large. Moreover, in the presence of a
transverse magnetic field, we show that the effective hopping
integrals acquire an unusual dependence on the magnetic field, through
Aharonov-Bohm interference, in addition to significant retardation
effects. The effect of the Aharonov-Bohm interference on the cyclotron
frequency (for small magnetic fields) and the 
Hofstadter butterfly (for large magnetic fields) is analyzed.
\end{abstract}

\pacs{72.10.-d,71.10.-w.,72.10.Di}

\maketitle

\section{Introduction}
Ever since the discovery  of high-temperature superconductivity in
cuprates,\cite{BM86} understanding the motion of a hole in an otherwise
half-filled CuO$_2$ layer has been a major challenge in condensed
matter physics.\cite{Be09} If, as most customary, one models this system with a
one-band Hubbard model,\cite{Hu63,*Ka63} then in the limit of a 
large on-site repulsion
$U$ the half-filled case maps onto an antiferromagnetic (AFM)
Heisenberg Hamiltonian with spin-exchange coupling
$J=4t^2/U$, where $t$ is the
nearest-neighbor (NN) hopping.\cite{GJR87} In this limit, then, 
the problem reduces to understanding the motion of a hole 
in a two-dimensional (2D) AFM background.\cite{Tr88,KLR89,MH91a}

A major reason for the difficulty in dealing with this question
is that despite being long-range ordered at $T=0$, the 
undoped AFM background
has a very complicated wavefunction due to spin fluctuations, which
make it very unlike the semi-classical N\'eel state. A simple but more
realistic description that could be used for analytical purposes is
missing; as a result, progress has been made primarily through numerical
simulations.\cite{Daea90,*FWRB91,*Da94,LNW06,*Le08}

Here, we present an essentially exact numerical solution, and
an approximate but quite accurate analytical approach, 
for the much simpler issue  of a hole moving in an Ising AFM 
background,\cite{SS88,*EBS90}
whose undoped wavefunction is the N\'eel state. The solutions have a
variational interpretation, and are shown to be accurate for a wide range of
parameters, of up to $t/J \approx 3$.

The generally accepted view, which probably explains why this problem 
has not been solved so far
(to the best of our knowledge),
is that a hole cannot really move in an Ising AFM, since as it hops
away from its initial location it re-shuffles the spins at the sites
it visits, creating a string of ``defects'' (misaligned spins). 
In dimensions higher than one, the energy cost of this string 
increases roughly linearly with its length, and 
as a result the hole is forced to stay in the
vicinity of its original position, i.e. its effective mass is
infinite.\cite{BR70} In this view, spin fluctuations which 
effectively remove (or ``heal'') pieces of this string of defects 
are needed to free the hole
and allow it to acquire a finite effective mass. Of course, such spin
fluctuations are absent in an Ising AFM.

That the hole is not truly localized even in an Ising AFM has been
known ever since Trugman pointed out\cite{Tr88} that by going
twice around a closed loop, the hole can actually  acquire a
finite effective mass.  We will return to this in more detail below, but
the main idea is that while the first circuit along the closed loop
creates the usual chain of defects, the second reshuffle of the spins
during the second circuit removes all these defects, but also ends
with the hole at a different location than the original site. By
repeating such processes, the hole can therefore move anywhere on its
original sublattice.

Even though Trugman
started from a N\'eel background and identified this
mechanism for generating a finite quasiparticle 
mass  in the absence of spin fluctuations,
he included spin-fluctuations in 
his calculation by considering
the full Heisenberg AFM Hamiltonian when determine 
the effective mass of the hole.\cite{Tr88} The issue of the hole
mass in a purely Ising AFM thus remained unanswered.

Here we carry out this calculation, and show that the hole  is fairly
mobile if $t\sim J$. Moreover, if a transverse magnetic field
is turned on, due to Aharonov-Bohm interference of the Peierls phases
associated with these closed loops,
the effective hoppings acquire a magnetic field dependence over and
above the usual Peierls phases, with interesting consequences. Our
results reveal
that even this seemingly simple problem
is actually very interesting and leads to rather non-trivial results. 

The paper is organized as follows. In Sec.~II, we specify the
model and our notation. In Sec.~III, we describe the
analytical calculation in the smallest variational subspace where the
hole acquires mass, both with and without magnetic field. 
We then explain the generalization for the
numerical calculation. Section~IV contains 
our results, and a summary and final conclusions appear in Sec.~V.

\section{Model}

Consider a spin-$\frac{1}{2}$ Ising AFM on a square
lattice
%%%%%%%%%%%%%%%%%%%%%%%%%%%%%% EQUATION %%%%%%%%%%%%%%%%%%%%%%%%%%%%%%
\begin{equation}
\EqLabel{m1}
{\cal H}_{\rm AFM} = J \sum_{\langle i, j\rangle}^{} \Big[S_{i,z}
  S_{j,z}-\frac{1}{4}\Big]  = \bar{\jmath} \sum_{\langle i, j\rangle}^{}
\left(\sigma_{i,z} 
\sigma_{j,z}  -1\right)\,,
\end{equation}
%%%%%%%%%%%%%%%%%%%%%%%%%%%%%%%%%%%%%%%%%%%%%%%%%%%%%%%%%%%%%%%%%%%%%%
where $\sigma_z$ is the Pauli matrix, and $\bj=J/4$. 
The ground state is
the classical N\'eel state, with  all spins on sublattice $A$  up, and
all spins on sublattice $B$  down,
\begin{equation}
\EqLabel{GS}
|{\rm GS}\rangle = \prod_{i\in A}^{} c^\dagger_{i, \uparrow} \prod_{j\in
  B}^{} c^\dagger_{j, \downarrow}|0\rangle \,, 
\end{equation}
where $c_{i,\sigma}$ annihilates an electron at site $i$, 
and for convenience we shifted the energy so that ${\cal H}_{\rm AFM}
|{\rm GS}\rangle = 0$. 

We would like to study the dynamics of a single hole in this
system, within the approximation that only NN hopping of
electrons is possible, however no-double occupancy is allowed. As  a
result, at each site we have either the hole, or a spin. Moreover, the
spin can only be in its proper orientation (consistent with the
sublattice its site belongs to) or flipped ({i.e.}, a frozen
magnon-like ``defect'' is 
created at this site).

In the following we will only keep track of the
location of the hole and of the flipped spins. To simplify the 
notation, we introduce a 
``defect'' creation operator
%%%%%%%%%%%%%%%%%%%%%%%%%%%%%% EQUATION %%%%%%%%%%%%%%%%%%%%%%%%%%%%%%
\begin{equation}
\EqLabel{m2}
d_i^{\dagger} =
\left\{
\begin{array}[c]{c}
\sigma_i^{-}, \mbox{ if } i\in A \\
\sigma_i^{+}, \mbox{ if } i\in B  \\
\end{array}
\right.\,,
\end{equation}
%%%%%%%%%%%%%%%%%%%%%%%%%%%%%%%%%%%%%%%%%%%%%%%%%%%%%%%%%%%%%%%%%%%%%%
where $\sigma^{\pm}$ are the raising/lowering Pauli matrices, and
hole creation operators  
%%%%%%%%%%%%%%%%%%%%%%%%%%%%%% EQUATION %%%%%%%%%%%%%%%%%%%%%%%%%%%%%%
\begin{equation}
\EqLabel{m3}
h_i^\dagger=
\left\{
\begin{array}[c]{c}
c_{i,\uparrow}, \mbox{ if } i\in A \\
c_{i,\downarrow}, \mbox{ if } i\in B  \\
\end{array}
\right.
\end{equation}
%%%%%%%%%%%%%%%%%%%%%%%%%%%%%%%%%%%%%%%%%%%%%%%%%%%%%%%%%%%%%%%%%%%%%%
for a  hole at the site $i$. With this notation, for example $h_i^\dagger
d_j^{\dagger}|{\rm GS}\rangle$ means that the hole is at site $i$ and the
spin at site $j\ne i$ is flipped; all other sites have their spins
in the proper  N\'eel configuration.

Consider now the motion of the hole, with the no-double occupancy condition
enforced. If the hole is at site $i$, the only possibility is for an
electron from one of its four NN sites $j$ to hop into $i$,
thus moving the hole to site $j$. If the spin of the electron at $j$ was properly
oriented, when it moves to $i$ it has the wrong orientation; in other
words, a ``defect'' is created at $i$ when the hole hops from
$i\rightarrow j$. On the other hand, if there was a ``defect'' at $j$
to begin with, when the electron moves to $i$ it will be properly
oriented, and therefore the ``defect'' at $j$ is removed as the hole hops
from $i\rightarrow j$.

Thus, the Hamiltonian that describes the dynamics of the hole in the
2D Ising AFM is 
%%%%%%%%%%%%%%%%%%%%%%%%%%%%%% EQUATION %%%%%%%%%%%%%%%%%%%%%%%%%%%%%%
\begin{equation}
\EqLabel{m4}
{\cal H} = P \Big[- \sum_{\langle i, j\rangle }^{} t_{ji} h_j^\dagger h_i
  \left( d_i^\dagger + d_j\right) + {\rm H.c.} 
\Big] P +  {\cal H}_{\rm AFM}.
\end{equation}
%%%%%%%%%%%%%%%%%%%%%%%%%%%%%%%%%%%%%%%%%%%%%%%%%%%%%%%%%%%%%%%%%%%%%%
The projector $P$ enforces
the no-double occupancy constraint,
as well as the condition that each site has either the hole, or a spin:
$h_i^\dagger h_i + d_i^\dagger d_i + d_i d_i^\dagger = 1$. In the presence of a
uniform transverse magnetic field $B$, 
the hopping integrals $t_{ij}$ include the proper 
Peierls factors (see below). 

This Hamiltonian is similar to the Edwards model,\cite{Ed06,*AEF07}
however, unlike that model it enforces (i) the fact that there can be
no magnons at the site where the hole is, {i.e.} states like $h_i^\dagger
d_i^{\dagger}|{\rm GS}\rangle$ are forbidden; (ii) the fact that there can be
at most one spin-flip per site, {i.e.} states like $h_i^\dagger
(d_j^{\dagger})^n|{\rm GS}\rangle, n\ge 2$ are forbidden. Finally, (iii) unlike the
Edwards model, where the energy of the defects was assumed to be
described by  $\Omega \sum_{i}^{} d_i^\dagger
d_i$, here we use ${\cal H}_{\rm AFM}$ to calculate the true cost for
creating spin flips -- two neighboring defects cost less energy
than two farther-apart ones, because they only disrupt seven AFM bonds,
not eight. Such corrections are less important than
is imposing (i) and (ii), but keeping track 
of the proper exchange energies is simple enough and we do so.

Note that a description of a Heisenberg AFM would require the addition of
terms  $\propto 2\bj \sum_{\langle i, j\rangle
}^{}(d_id_j+d^\dagger_id^\dagger_j ) $, describing the 
XY spin-exchange interaction
leading to spin fluctuations. However, since the N\'eel
ground-state $|{\rm GS}\rangle$ is not a good approximation for the
undoped ground state of this model, it is hard to quantify the meaning
of adding these terms.
 Spin fluctuations
could also be introduced  with a small magnetic
field parallel to the $x$-axis, leading to a term $\propto \sum_{i}^{}
( d_i^\dagger + d_i)$, as considered in the Edwards model.\cite{EEAF10} 
As shown there, such terms do significantly lower the effective mass of the
hole, but  we ignore them here.

\section{Formalism}

\subsection{Hole propagation when $B=0$}

We begin by considering the case where no transverse magnetic field is
applied. In this case, the unit cell contains two sites (one from
sublattice $A$, one from sublattice $B$) 
and the corresponding magnetic Brillouin zone is
a square rotated by $\pi/4$ (see below). For each $\mb{k}$
  in this Brillouin zone, we define the plane wave
%%%%%%%%%%%%%%%%%%%%%%%%%%%%%% EQUATION %%%%%%%%%%%%%%%%%%%%%%%%%%%%%%
\begin{equation}
\EqLabel{m5}
h^\dagger_{\mb{k}} = {1\over \sqrt{\bar{N}}} \sum_{i \in A}^{}
e^{i\mb{k}\cdot \mb{R}_i} c_{i\uparrow} \,,
\end{equation}
%%%%%%%%%%%%%%%%%%%%%%%%%%%%%%%%%%%%%%%%%%%%%%%%%%%%%%%%%%%%%%%%%%%%%%
and want to calculate the Green's function
%%%%%%%%%%%%%%%%%%%%%%%%%%%%%% EQUATION %%%%%%%%%%%%%%%%%%%%%%%%%%%%%%
\begin{equation}
\EqLabel{m6}
G(\mb{k},\omega) = \langle {\rm GS}| h_{\mb k} \hat{G}(\omega)
h^\dagger_{\mb k}|{\rm GS}\rangle \,,
\end{equation}
%%%%%%%%%%%%%%%%%%%%%%%%%%%%%%%%%%%%%%%%%%%%%%%%%%%%%%%%%%%%%%%%%%%%%%
where 
\begin{equation}
\hat{G}(\omega)= \frac{1}{\omega + i \eta - {\cal H}}
\end{equation}
is the resolvent associated with this Hamiltonian, $\eta>0$ is
infinitesimally small, and we set $\hbar=1$. $\bar{N}$ denotes 
the number of unit cells, or the number of sites in each sublattice, 
and is taken to infinity. 

Besides producing the single-hole spectrum from its poles, the spectral
function $A(\mb{k}, \omega) = -{1\over \pi} \mbox{Im} G({\mb k},
\omega)$ is the quantity that would be measured by spin-polarized
ARPES, assuming that a (photo)electron with spin-up is removed from
the system. Of course, there is a second set of wavefunctions
consisting of plane waves involving the sites on sublattice $B$; their
associated Green's function would be related to ARPES measured when a
spin-down electron was ejected from the system. Obviously, the
spectral weights are identical for the two cases.

As discussed, as the hole hops away from its initial site,
it creates a string of defects  as it reshuffles the spins on
its path. These defects increase the energy of the state roughly
linearly with the length of the string, because of the broken AFM
bonds. In the absence of spin fluctuations, there are only two ways to
remove this costly string of misaligned spins. One is for the hole to
retrace its path, removing all the defects -- in this case it ends up
at the original site it started from.\cite{BR70} Such terms can only 
renormalize the overall energy, but do not generate dispersion.
The second possibility is for the hole to go repeatedly around closed
loops: on the first circuit a string of defects is
created, while on the second circuit the defects are removed. 
These loops generate effective second and third NN hopping terms, 
as counted in the original square lattice.

\begin{figure}[t]
\includegraphics[width=\columnwidth]{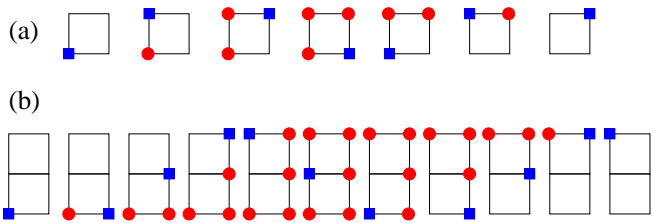}
\caption{(Color online) Shortest sequences of repeated hops
  around closed loops 
  leading to effective 2nd NN (top panel) and 3rd NN (bottom panel)
  hopping. The hole is marked by a blue square, and the defects (spin
  flips) it creates or annihilates as it moves are marked by red circles.
\label{fig1}}
\end{figure}

In Fig.~\ref{fig1} we show examples of the shortest sequences of
events that result in generating such 2nd (a) and 3rd (b) NN
hoppings.\cite{Tr88} One can easily convince oneself that longer
closed loops can only contribute to one of these two effective
hoppings, since the final location of the hole is within two hops of
the original one. Effective NN hopping is impossible, because of spin
conservation: if the hole is on the other sublattice, there must be an
odd number of magnons around.

In perturbational terms, for $t \ll J=4\bj$, the sequence of hoppings
depicted in Fig.~\ref{fig1}(a) results in 2nd NN hopping $t_2 =
t^6/(6^2 \cdot 10^2\cdot 12 \bj^5)$, since the initial and final
states, which are both in the no-defect manifold, are connected
through 6 hopping processes, and the five intermediary states have
energies higher by $6\bj, 10\bj, 12\bj, 10\bj, 6\bj$ respectively.
Similarly, Fig.~\ref{fig1}(b) gives a contribution to 3rd NN hopping
$t_3 = t^{10}/(6^2 \cdot 10^2\cdot 14^2\cdot 18^3 \bj^9)$, as the
longer strings of defects are more costly.

Note that clockwise hopping of the hole around the same loop as shown
in Fig.~\ref{fig1}(b) would generate a higher-order contribution to
the 2nd NN hopping $t_2'\approx t_3$ because some of the intermediary
states have slightly different energies. This allows us to estimate
the range where perturbation theory is valid. If we ask that $t_2'/t_2
\le 0.1$, so that this correction from 5-defect processes to the
3-defect result is small, we find $t\le 8.4\bj= 2.1J$. In other words,
because of the relatively long and thus energetically costly loops of
flipped spins involved in generating the effective hoppings,
perturbation theory extends up to rather large values $t \le 2J$. For
comparison, in cuprates it is believed that $t\sim J/3$, so they would
be just outside the perturbational range (for these values, $t_2'\sim
0.4 t_2$). Of course, the effects of spin fluctuations cannot be
ignored in the cuprates.

Clearly, then, for $t\le 2J$ the dominant contribution to the hole's dynamics
comes from the processes of Fig.~\ref{fig1}(a). We therefore begin by
calculating $G(\mb{k},\omega)$ in a variational approach, by only
allowing such short loops (with up to 3 defects on 3 corners on a
square plaquette) to be generated. Longer loops are more expensive and
the probability to generate them should be lower. Of course, they can
be included in a numerical calculation, but this will lead to only
quantitative, not qualitative  changes (see below).

The method we use is similar to that used in Refs.~\onlinecite{EEAF10,AEF10} 
for the 1D, and in Ref.~\onlinecite{BF10} for the 2D Edwards model.
Within this variational approximation, we generate the equations of
motion for the propagator by using repeatedly the Dyson identity
\begin{equation}
\hat{G}(\omega) = \hat{G}_0(\omega) +  \hat{G}(\omega) V
\hat{G}_0(\omega)\,, 
\end{equation}
where $\hat{G}_0(\omega)$ is the resolvent for
${\cal H}_0= {\cal H}_{\rm AFM}$, while $V$ is the first term in
Eq.~(\ref{m4}), describing the hopping of the hole.
We then find
%%%%%%%%%%%%%%%%%%%%%%%%%%%%%% EQUATION %%%%%%%%%%%%%%%%%%%%%%%%%%%%%%
\begin{equation}
\EqLabel{m7}
G(\mb{k},\omega) = G_0(\omega- 4\bj)\Big[1- t \sum_{\bdelta}^{}
  F_1(\mb{k},\omega, {\bdelta})\Big]\,, 
\end{equation}
%%%%%%%%%%%%%%%%%%%%%%%%%%%%%%%%%%%%%%%%%%%%%%%%%%%%%%%%%%%%%%%%%%%%%%
where $G_0(\omega)=1/(\omega+i\eta)$, $4\bj$ is the cost of having only the
hole in the system (first state in Fig.~\ref{fig1}(a)), and
$\bdelta$
points to any 
of the four NN sites, i.e. ${\bm \bdelta} \in \{(\pm1, 0),
  (0, \pm1)\}$ for a 
lattice constant $a=1$. 
The new propagators are 
%%%%%%%%%%%%%%%%%%%%%%%%%%%%%% EQUATION %%%%%%%%%%%%%%%%%%%%%%%%%%%%%%
\begin{equation}
\EqLabel{m81}
%\nonumber
F_1(\mb{k},\omega, \bdelta) = {1\over \sqrt{N}}
\sum_{i \in A}^{} 
e^{i\mb{k}\cdot \mb{R}_i} \langle {\rm GS}| h_{\mb k} \hat{G}(\omega)
d^\dagger_ih^\dagger_{i+\bdelta}|{\rm GS}\rangle \,,
\end{equation}
%%%%%%%%%%%%%%%%%%%%%%%%%%%%%%%%%%%%%%%%%%%%%%%%%%%%%%%%%%%%%%%%%%%%%%
and are related to the amplitude of probability to propagate between a
no-defect state and a 1-defect state such as the second state shown
in Fig.~\ref{fig1}(a), with a defect at the original site. In the
following, for simplicity we will denote $F_1(\mb{k},\omega,
\bdelta)$ as $F_1(\bdelta)$.

An equation of motion for $F_1$  can now be generated. The
hopping $V$ can lead to three different outcomes: either the hole hops
back to site $i$, removing the defect; or it hops by another
$\bdelta$, creating a linear string of the type
$d^\dagger_id^\dagger_{i+\bdelta}h^\dagger_{i+2\bdelta}|{\rm GS}\rangle$; or
it hops by one of the two $\bdelta' \perp \bdelta$, leading to a state
such as the 
3rd state  shown
in Fig.~\ref{fig1}(a). Only the first and last outcomes are allowed
within our variational calculation, leading to\begin{widetext}
%%%%%%%%%%%%%%%%%%%%%%%%%%%%%% EQUATION %%%%%%%%%%%%%%%%%%%%%%%%%%%%%%
\begin{equation}
\EqLabel{m82}
%\nonumber
F_1(\bdelta_1) = - t G_0(\omega-10\bj) \left[G(\mb{k},\omega) +
  \sum_{\bdelta_2 \perp \bdelta_1}^{} F_2(\bdelta_1, \bdelta_2)\right]\,,
\end{equation}
%%%%%%%%%%%%%%%%%%%%%%%%%%%%%%%%%%%%%%%%%%%%%%%%%%%%%%%%%%%%%%%%%%%%%%
where
%%%%%%%%%%%%%%%%%%%%%%%%%%%%%% EQUATION %%%%%%%%%%%%%%%%%%%%%%%%%%%%%%
\begin{equation}
\EqLabel{m9}
%\nonumber
F_2(\bdelta_1, \bdelta_2) = 
\sum_{i \in A}^{} {e^{i\mb{k}\cdot \mb{R}_i}\over \sqrt{N}}
 \langle {\rm GS}| h_{\mb k} \hat{G}(\omega)
d^\dagger_id^\dagger_{i+\bdelta_1}h^\dagger_{i+\bdelta_1+ \bdelta_2}|{\rm GS}\rangle
\end{equation}
%%%%%%%%%%%%%%%%%%%%%%%%%%%%%%%%%%%%%%%%%%%%%%%%%%%%%%%%%%%%%%%%%%%%%%
describes generalized Green's functions associated with 2-defect
states. 
Its equation of motion, within our variational space, is 
%%%%%%%%%%%%%%%%%%%%%%%%%%%%%% EQUATION %%%%%%%%%%%%%%%%%%%%%%%%%%%%%%
\begin{equation}
\EqLabel{m10}
F_2(\bdelta_1, \bdelta_2) = -t G_0(\omega-14\bj)\left[F_1(\bdelta_1) +
  F_3(\bdelta_1,\bdelta_2, -\bdelta_1) \right]\,,
\end{equation}
%%%%%%%%%%%%%%%%%%%%%%%%%%%%%%%%%%%%%%%%%%%%%%%%%%%%%%%%%%%%%%%%%%%%%%
where
%%%%%%%%%%%%%%%%%%%%%%%%%%%%%% EQUATION %%%%%%%%%%%%%%%%%%%%%%%%%%%%%%
\begin{equation}
\EqLabel{m11}
F_3(\bdelta_1, \bdelta_2, \bdelta_3) = \sum_{i \in A}^{} {e^{i\mb{k}\cdot
	\mb{R}_i}\over \sqrt{N}} 
 \langle {\rm GS}| h_{\mb k} \hat{G}(\omega)
d^\dagger_id^\dagger_{i+\bdelta_1}d^\dagger_{i+\bdelta_1+
 \bdelta_2}h^\dagger_{i+\bdelta_1+ \bdelta_2+\bdelta_3}|{\rm GS}\rangle
\end{equation}
%%%%%%%%%%%%%%%%%%%%%%%%%%%%%%%%%%%%%%%%%%%%%%%%%%%%%%%%%%%%%%%%%%%%%%
%\end{widetext}
describes  states with a string of three consecutive defects. Only states
such as shown in the middle panel of Fig.~\ref{fig1}(a) are in our
variational space, hence the unique $F_3$ term in the equation of
$F_2$. Finally, the equation for $F_3$ is now connected to two
possible 2-defect states. The hole could either hop back, linking 
to $F_2(\bdelta_1, \bdelta_2)$, or it could complete the closed loop by hopping
to site $i$ and removing the defect that was there, resulting in 
%\begin{widetext}
%%%%%%%%%%%%%%%%%%%%%%%%%%%%%% EQUATION %%%%%%%%%%%%%%%%%%%%%%%%%%%%%%
\begin{equation}
\EqLabel{m11x}
\sum_{i \in A}^{} {e^{i\mb{k}\cdot \mb{R}_i}\over \sqrt{N}}
 \langle {\rm GS}| h_{\mb k} \hat{G}(\omega)
h^\dagger_id^\dagger_{i+\bdelta_1}d^\dagger_{i+\bdelta_1+
 \bdelta_2}|{\rm GS}\rangle = e^{-i \mb{k} (\bdelta_1+  \bdelta_2)}
  F_2(-\bdelta_2, -\bdelta_1)\,.
\end{equation}
%%%%%%%%%%%%%%%%%%%%%%%%%%%%%%%%%%%%%%%%%%%%%%%%%%%%%%%%%%%%%%%%%%%%%%
Eq. (\ref{m11x})  follows after a translation by
$-(\bdelta_1+\bdelta_2)$, which maintains the site $i$ on the same
original sublattice $A$, since $\bdelta_2 \perp \bdelta_1$.
Of course,  hopping also links $F_3$ to two generalized propagators
$F_4$ with 4 defects, however those are ignored in  this 3-defect variational
space. Thus, we find:
%%%%%%%%%%%%%%%%%%%%%%%%%%%%%% EQUATION %%%%%%%%%%%%%%%%%%%%%%%%%%%%%%
\begin{equation}
\EqLabel{m12}
F_3(\bdelta_1, \bdelta_2, -\bdelta_1) = -t G_0(\omega-16\bj)
\left[F_2(\bdelta_1, \bdelta_2) + e^{-i \mb{k} \cdot (\bdelta_1+  \bdelta_2)}
  F_2(-\bdelta_2, -\bdelta_1)\right]\,.
\end{equation}
%%%%%%%%%%%%%%%%%%%%%%%%%%%%%%%%%%%%%%%%%%%%%%%%%%%%%%%%%%%%%%%%%%%%%%
In other words we have a closed system of linear equations for $G$ and
the various allowed $F_1, F_2, F_3$ functions.

This system can
be solved analytically. The final result is
%%%%%%%%%%%%%%%%%%%%%%%%%%%%%% EQUATION %%%%%%%%%%%%%%%%%%%%%%%%%%%%%%
\begin{equation}
\EqLabel{m13}
G(\mb{k},\omega) = \frac{1}{\omega+i\eta -\epsilon_0(\omega) + 4
  t_2(\omega) \cos k_x \cos 
k_y + 4 t_3(\omega) \,[\cos(2k_x) + \cos(2k_y)]}\,,
\end{equation}
\end{widetext}
%%%%%%%%%%%%%%%%%%%%%%%%%%%%%%%%%%%%%%%%%%%%%%%%%%%%%%%%%%%%%%%%%%%%%%
where
%%%%%%%%%%%%%%%%%%%%%%%%%%%%%% EQUATION %%%%%%%%%%%%%%%%%%%%%%%%%%%%%%
\begin{equation}
\EqLabel{m15}
\epsilon_0 (\omega) = 4\bj + 4t^2 G_0({\tilde{\tilde
\omega}}) - 4 t_3\,,
\end{equation}
%%%%%%%%%%%%%%%%%%%%%%%%%%%%%%%%%%%%%%%%%%%%%%%%%%%%%%%%%%%%%%%%%%%%%%
%%%%%%%%%%%%%%%%%%%%%%%%%%%%%% EQUATION %%%%%%%%%%%%%%%%%%%%%%%%%%%%%%
\begin{equation}
\EqLabel{m14}
t_2(\omega) = - \frac{2 t^2 G_0({\tilde{\tilde
\omega}})\beta(\omega)}{1- 4 \beta^2(\omega)}\,,
\end{equation}
%%%%%%%%%%%%%%%%%%%%%%%%%%%%%%%%%%%%%%%%%%%%%%%%%%%%%%%%%%%%%%%%%%%%%%
%%%%%%%%%%%%%%%%%%%%%%%%%%%%%% EQUATION %%%%%%%%%%%%%%%%%%%%%%%%%%%%%%
\begin{equation}
\EqLabel{m14b}
t_3(\omega) = \beta(\omega) t_2(\omega)\,.
\end{equation}
%%%%%%%%%%%%%%%%%%%%%%%%%%%%%%%%%%%%%%%%%%%%%%%%%%%%%%%%%%%%%%%%%%%%%%

The functions that appear in these definitions are:
%%%%%%%%%%%%%%%%%%%%%%%%%%%%%% EQUATION %%%%%%%%%%%%%%%%%%%%%%%%%%%%%%
\begin{equation}
\EqLabel{m16}
\beta(\omega) = \frac{t^2 G_0({\tilde \omega}) G_0({\tilde{\tilde
\omega}}) \gamma(\omega)}{1- \gamma^2(\omega)}\,,
\end{equation}
%%%%%%%%%%%%%%%%%%%%%%%%%%%%%%%%%%%%%%%%%%%%%%%%%%%%%%%%%%%%%%%%%%%%%%
%%%%%%%%%%%%%%%%%%%%%%%%%%%%%% EQUATION %%%%%%%%%%%%%%%%%%%%%%%%%%%%%%
\begin{equation}
\EqLabel{m17}
{\tilde{\tilde
\omega}} = \omega - 10\bj - \frac{2t^2 G_0({\tilde \omega})}{1-
  \gamma^2(\omega)}\,, 
\end{equation}
%%%%%%%%%%%%%%%%%%%%%%%%%%%%%%%%%%%%%%%%%%%%%%%%%%%%%%%%%%%%%%%%%%%%%%
%%%%%%%%%%%%%%%%%%%%%%%%%%%%%% EQUATION %%%%%%%%%%%%%%%%%%%%%%%%%%%%%%
\begin{equation}
\EqLabel{m18}
\gamma(\omega) = t^2 G_0({\tilde \omega})G_0(\omega-16 \bj),
\end{equation}
%%%%%%%%%%%%%%%%%%%%%%%%%%%%%%%%%%%%%%%%%%%%%%%%%%%%%%%%%%%%%%%%%%%%%%
and 
%%%%%%%%%%%%%%%%%%%%%%%%%%%%%% EQUATION %%%%%%%%%%%%%%%%%%%%%%%%%%%%%%
\begin{equation}
\EqLabel{m19}
{\tilde \omega}= \omega - 14\bj - t^2 G_0(\omega-16\bj)\,.
\end{equation}
%%%%%%%%%%%%%%%%%%%%%%%%%%%%%%%%%%%%%%%%%%%%%%%%%%%%%%%%%%%%%%%%%%%%%%

The Green's function of Eq.~(\ref{m13}) describes a free particle
on the $A$ sublattice, with an onsite energy $\epsilon_0$, and
hoppings $t_2$ and $t_3$ that correspond to NN and 2nd NN for sites on
this sublattice. Interestingly, all these effective quantities are
functions of $\omega$, in other words retardation effects are
explicitly taken into consideration in this variational
calculation. For $t\ll \bj$ we find that indeed $\epsilon_0 \approx
4\bj-4t^2/6\bj$, while $t_2 = t^6/(21600\bj^5)$. This is precisely twice the
value we estimated for Fig.~\ref{fig1}(a); the factor of 2 is due to 
contributions from both clockwise and 
anticlockwise loops.

\begin{figure}[t]
\includegraphics[width=\columnwidth]{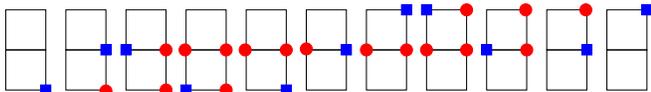}
\caption{(Color online) An effective  $t_3$ process generated with
  only 3-defect strings. 
\label{fig2}}
\end{figure}

The appearance of the $t_3$ term at this level of
the calculation is surprising at first, since configurations with 5
defects like
that appearing in Fig.~\ref{fig1}(b) are not allowed in this variational
space. The answer for how to achieve 
$t_3$ hopping with only 3-defect loops is shown in Fig.~\ref{fig2}:
essentially, instead of removing the last defect at the 6th hopping,
another 3-defect process is initiated. It is straightforward 
to check that in the perturbational limit, the ratio between the
expected values for this $t_3$ and the $t_2$ of Fig.~\ref{fig1}(a) is
of $t^4/(7200 \bj^4)$. This is precisely the ground-state value of 
$\beta(\omega=4\bj)$ in
this asymptotic limit, which further validates this statement. It is
also now clear that this strategy of starting a new loop when only a
single  defect is left can be repeated an arbitrary number
of times, and an infinite sequence of higher order contributions to
$t_2$ and $t_3$ can be 
generated this way. This explains the $1-4\beta^2(\omega)$
denominators in Eqs.~(\ref{m14}),(\ref{m14b}), showing that our
variational calculation sums contributions from these 3-defect processes to all
orders.

As argued previously, the contribution of 5-defect loops is negligible
if $t\le 2J=8\bj$. Interestingly, for such values $2\beta \sim
t^4/(3600\bj^4)$ varies from 0 to just above 1 (however, this ignores
the change in $\beta(\omega)$ because the ground-state energy moves
away from $\omega=4\bj$, as $t/J$ increases). In other words, the
denominators in Eqs.~(\ref{m14}), (\ref{m14b}) could become very
small, implying that significant hopping, and thus light effective
masses, may be possible in such systems even for rather small $t/J$
ratios.  

As a final comment, the fact that $\mb{k}$-dependence appears only in the
equation for $F_3$  confirms that the effective hopping terms are
due to completely circling around the closed loops. If we truncated
the variational space to 
include only 2-defect states, which would mean setting $F_3 \rightarrow 0$ in
Eq.~(\ref{m10}), the result would be just a renormalization of the
on-site energy but no dispersion.

\subsection{Hole propagation when $B\ne 0$}

Since the dispersion of the hole is due to Trugman loops, it is
interesting to consider what happens when a transverse magnetic field
$B$ is applied. Because of the large number of elementary hopping
processes involved in generating effective longer ranged hoppings, one
would expect the phases associated with these effective hoppings to be
different from the normal Peierls phase. Put it another way, one
would expect Aharonov-Bohm-like interference between clockwise and
counterclockwise contributions, over and above the usual Peierls
phases, and therefore a spectrum that is not simply that of a bare
particle with hoppings $t_2, t_3$ placed in a magnetic field.

Let $\Phi = Ba^2$ be the magnetic flux through the elementary square
plaquette. Because the dressed hole only lives on one sublattice
(sublattice $A$ in our calculation), the actual relevant flux is
$2Ba^2$, since the unit cell for the sublattice is doubled in size.

From elementary considerations of the Hofstadter butterfly,\cite{Ho76}
we know that we need to consider the cases when the ratio between
the magnetic flux and the elementary flux $\Phi_0=h/e$ is
$2Ba^2/\Phi_0 = p/n$, where $p$ and $n$ are mutually prime
integers. In this case, we expect the spectrum to split into $n$
magnetic subbands. Of course, then, the Brillouin zone is
  folded down $n$ times.  We are therefore interested in the spectrum
of the dressed hole when
\begin{equation}
\frac{\Phi}{ \Phi_0}= {p\over 2n}
\end{equation}
(for simplicity of notation, we also introduce $\phi = 2 \pi
  \Phi/\Phi_0 = \pi p/n$).

\begin{figure}[t]
\includegraphics[width=0.6\columnwidth]{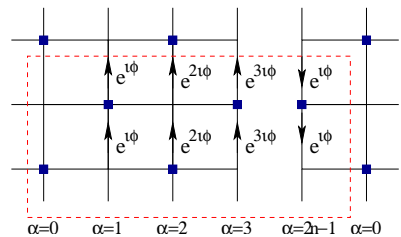}
\caption{(Color online) Magnetic unit cell in real space when $\phi =
  2\pi p/(2n)$, where $p$ and $n$ are mutually prime integers. The
  unit cell is marked by the red dashed line, and contains $2n$ sites,
  labeled by squares, on the $A$ sublattice. These are indexed by their
  corresponding $\alpha=0,...,2n-1$. The value of $\alpha$ also
  appears in the Peierls phases for hopping in the direction of the arrow.
\label{fig3}}
\end{figure}

We use the Landau gauge $\mb{A}({\mb r})= Bx \mb{e}_y$, in which case
only the hopping integrals in the $y$-direction acquire Peierls
phases. The magnetic unit cell will contain $2n$ sublattice $A$ sites,
and it is drawn, together with the Peierls phases, in
Fig.~\ref{fig3}. Each site in the unit cell is labeled by the index
$\alpha=0,\ldots,2n-1$ which establishes its location in the unit
cell, as well as the phase associated with hopping off this site in
the positive $y$-direction.

The Hamiltonian is identical to that of Eq.~(\ref{m1}), with the
proper phases included in the hoppings. Of course, we assume that the
magnetic field is not sufficiently large to overcome the AFM coupling
and favor a FM ground state. Also, we ignore the Zeeman term -- since
the spins are  reshuffled as the hole hops around, the total
$z$-axis spin $S_z= \sum_{i}^{} \sigma_{i,z}$ remains constant and
therefore the Zeeman contribution is an overall constant.

We now need to introduce $2n$ distinct plane waves, associated with
each site of the unit cell
%%%%%%%%%%%%%%%%%%%%%%%%%%%%%% EQUATION %%%%%%%%%%%%%%%%%%%%%%%%%%%%%%
\begin{equation}
\EqLabel{n1}
h^\dagger_{\alpha,\mb{k}} = \sqrt{{2n\over \bar{N}}} \sum_{i \in A_\alpha}^{}
e^{i\mb{k}\cdot \mb{R}_i} c_{i\uparrow}\,, 
\end{equation}
%%%%%%%%%%%%%%%%%%%%%%%%%%%%%%%%%%%%%%%%%%%%%%%%%%%%%%%%%%%%%%%%%%%%%%
where $A_\alpha$ collects all sites on sublattice $A$ with the same
value of $\alpha$. The procedure to generate the equations 
of motion for the propagator
%%%%%%%%%%%%%%%%%%%%%%%%%%%%%% EQUATION %%%%%%%%%%%%%%%%%%%%%%%%%%%%%%
\begin{equation}
\EqLabel{n2}
G_{\alpha'\alpha}(\mb{k},\omega) = \langle {\rm GS}| h_{\alpha',\mb k} \hat{G}(\omega)
h^\dagger_{\alpha,\mb k}|{\rm GS}\rangle
\end{equation}
%%%%%%%%%%%%%%%%%%%%%%%%%%%%%%%%%%%%%%%%%%%%%%%%%%%%%%%%%%%%%%%%%%%%%%
is just as before,
however now one has to take into 
consideration the Peierls phases. These can either be included in the
definition of the new $F_1, F_2, F_3,$ functions, or can be explicitly
pulled out. In either case, the overall structure of the system of
linear equations is very similar to that in the $B=0$ case,
and can be solved similarly. This allows us to  reduce it to a single
equation involving only $G$ functions, which reads:
\begin{widetext}
%%%%%%%%%%%%%%%%%%%%%%%%%%%%%% EQUATION %%%%%%%%%%%%%%%%%%%%%%%%%%%%%%
\begin{multline}
\EqLabel{n3}
G_{\alpha' \alpha}(\mb{k},\omega)  = G_0(\omega- \tilde{\epsilon}_0)
\left\{\delta_{\alpha'\alpha} - \tilde{t}_2 \left[2\cos( k_y -(\alpha -
	\tfrac{1}{2})\phi) e^{ik_x} G_{\alpha',\alpha-1}(\mb{k},\omega)  +
	2\cos( k_y -(\alpha +	\tfrac{1}{2})\phi) e^{-ik_x}
	G_{\alpha',\alpha+1}(\mb{k},\omega)  \right] 
  \right. \\
\left. - \tilde{t}_3 \left[ e^{2ik_x} G_{\alpha', \alpha-2}(\mb{k},\omega)  + 2
  \cos(2k_y-2\alpha\phi) G_{\alpha'\alpha}(\mb{k},\omega)  + e^{-2ik_x}
  G_{\alpha',\alpha+2}(\mb{k},\omega) \right]\right\}\,.
\end{multline}
%%%%%%%%%%%%%%%%%%%%%%%%%%%%%%%%%%%%%%%%%%%%%%%%%%%%%%%%%%%%%%%%%%%%%%
Here we used the short-hand notations:
%%%%%%%%%%%%%%%%%%%%%%%%%%%%%% EQUATION %%%%%%%%%%%%%%%%%%%%%%%%%%%%%%
\begin{equation}
\EqLabel{n4}
\tilde{\epsilon}_0 \equiv \epsilon_0 (\phi,\omega) = 4\bj +  \frac{4 t^2
  G_0({\tilde{\tilde 
\omega}})(1-2\beta^2(\omega))}{1- 4 \beta^2(\omega)+ 4
  \beta^4(\omega)  \sin^2(2\phi)}\,,
\end{equation}
%%%%%%%%%%%%%%%%%%%%%%%%%%%%%%%%%%%%%%%%%%%%%%%%%%%%%%%%%%%%%%%%%%%%%%
%%%%%%%%%%%%%%%%%%%%%%%%%%%%%% EQUATION %%%%%%%%%%%%%%%%%%%%%%%%%%%%%%
\begin{equation}
\EqLabel{n5}
\tilde{t}_2 \equiv t_2(\phi,\omega) = - \frac{2 t^2 G_0({\tilde{\tilde
\omega}})\beta(\omega)\left[(1-\beta^2(\omega))\cos{3\phi\over2} +
	\beta^2(\omega) \cos{5\phi\over2}\right]}{1- 4 \beta^2(\omega)+ 4
  \beta^4(\omega)  \sin^2(2\phi)}\,,
\end{equation}
%%%%%%%%%%%%%%%%%%%%%%%%%%%%%%%%%%%%%%%%%%%%%%%%%%%%%%%%%%%%%%%%%%%%%%
and
%%%%%%%%%%%%%%%%%%%%%%%%%%%%%% EQUATION %%%%%%%%%%%%%%%%%%%%%%%%%%%%%%
\begin{equation}
\EqLabel{n6}
\tilde{t}_3 \equiv t_3(\phi,\omega) = - \frac{2 t^2 G_0({\tilde{\tilde
\omega}})\beta^2(\omega)\cos(2\phi)}{1- 4 \beta^2(\omega)+ 4
  \beta^4(\omega)  \sin^2(2\phi)}\,.
\end{equation}
%%%%%%%%%%%%%%%%%%%%%%%%%%%%%%%%%%%%%%%%%%%%%%%%%%%%%%%%%%%%%%%%%%%%%%
\end{widetext}

Remarkably, Eq.~(\ref{n3}) is identical to the equation of motion one
would obtain for the propagator associated with a Hamiltonian on the
sublattice $A$ only, 
with an on-site energy 
$\tilde{\epsilon}_0$ and NN and 2nd NN hopping (as defined on the
sublattice) given by $\tilde{t}_2, \tilde{t}_3$, plus the proper
Peierls phases for the applied
magnetic field. The terms in the parenthesis multiplied by
$\tilde{t}_2$ come from NN hopping by $\pm \delta_x \pm \delta_y$, with
the proper Peierls phases accounted for in the phases of the cosine
functions; similarly, the terms in the 
parenthesis multiplied by  $\tilde{t}_3$ come from next-NN hopping 
by $\pm
2 \delta_x, \pm 2\delta_y$. Only the $ \pm 2\delta_y$ hopping
accumulates a Peierls phase, in this case, hence the cosine term appears
only for that term.

It follows that $\tilde{\epsilon}_0, \tilde{t}_2$ and $ \tilde{t}_3$
are the effective parameters describing the new quasiparticle (the
dressed hole in the N\'eel AFM  
background plus magnetic field). They 
depend strongly on the 
magnetic field, besides the $\omega$-dependence due to retardation
effects. As expected, they  have the correct values when $\phi\rightarrow 0$,
as given by Eqs.~(\ref{m15})-(\ref{m14b}). 

The $\cos{3\phi\over 2}$ and $\cos{5\phi \over 2}$ terms in
$\tilde{t}_2$, respectively $\cos{2\phi}$ term in $\tilde{t}_3$, are
due to Aharonov-Bohm interference between clockwise and
counterclockwise loops. This can be checked directly in the asymptotic
limit $t\ll J$ by counting the accumulated Peierls phases for the
processes from Figs.~\ref{fig1}(a) and~\ref{fig2}. Because of these,
we expect the resulting Hofstadter butterfly to have an unusual
periodicity.

That is, we still expect the appearance of $n$ bands when
$2\Phi/\Phi_0=p/n$, due to interference between the Peierls
phases. However, additional dependence of the hoppings on $\phi$
because of the Aharonov-Bohm interference will further modulate the
overall bandwidth with a period $2\pi$ in $\phi$ (the band structure
is invariant if only $t_2$ changes its sign). In contrast,
  normally, ({i.e.} for constant $\epsilon_0, t_2, t_3$) the spectrum
  is an even function of $\phi$ with period $\pi$, because if the
  magnetic flux through the unit cell ($2\Phi$, in this case),
  increases by $\Phi_0$, the spectrum stays unchanged.  Thus, we
  expect a quite different-looking Hofstadter butterfly for the
  dressed hole. In particular, since at least for small $t/J$ we
  expect $t_3 \ll t_2$, the spectrum should become extremely narrow
  every time $\tilde{t}_2\rightarrow 0$ because of destructive
  Aharonov-Bohm interference.

\subsection{Numerical calculation method }

The numerical calculation that we describe in this section 
has been carried out only for the case $B=0$. This is a
straightforward generalization of the 3-defect variational calculation
discussed above. The idea is to systematically increase the size of
the variational space, to see whether addition of configurations with
longer loops of defects changes results significantly. If it does not,
then we know that the calculation is converged and therefore
the results are essentially exact.

\begin{figure}[t]
\includegraphics[width=0.8\columnwidth]{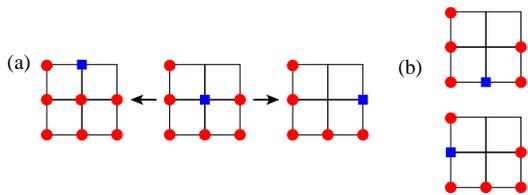}
\caption{(Color online) Examples of discarded configurations: (a) the
  central 6-defect configuration is only allowed to be linked back to
  the 5-defect configuration that generated it, shown to its right, or to the
  7-defect configuration shown to its left. The other two
  ``self-crossing'' 5-defect configurations shown in (b) are not kept in the
  variational space.
\label{fig4}}
\end{figure}

We use an index $N$ to characterize the maximum number of defects
allowed within the variational subspace. The equations of motion are
generated as before, starting from just the hole and keeping all configurations
linked to by hole hopping, up to those involving $N$ defects. The exception
is for longer chains with $N\ge 4$ defects, where we do not allow the
chain to ``self-cross'' itself. For instance, for the 6-defect chain
sketched in Fig.~\ref{fig4}, the hole is not allowed to hop either
left or down. Either of these processes would result in a
configuration with the hole
having two strings of defects attached to it. Trying to remove all
these defects  involves a very
complicated sequence of Trugman loops, and is therefore expected to
have a very small 
contribution to $t_2$ and $t_3$. The other role of such
configurations is to renormalize energies, but as we will see in the
next section, we can argue that their effect must be quite
negligible.

This is why for a configuration like that of Fig.~\ref{fig4}, we allow
the hole only to hop back (towards right) to the $F_5$ configuration
from which it derived; or to hop up to generate a $F_7$
configuration. In its turn, hopping to the left from this $F_7$
configuration will close the loop and give a new contribution to the
effective $t_2$. New possible closed loops (and therefore additional
contributions to $t_2,t_3$) appear for $N=3,5,7,\ldots$ 

Once all the allowed configurations have been generated for a given
$N$, the resulting (very sparse) linear system comprising their
equations of motion is solved numerically using
PARDISO.\cite{SG04,*SG06} We show results with up to $N=7$, in the
following, because here there are already over ten thousand allowed
configurations, and, as we will see, this is sufficiently large for
convergence to be achieved up to a quite high $t/J$ ratio.

Finally, note that in the numerical calculation, $N=3$  contains many more
configurations than in the $N=3$  analytical calculation, because now
all possible 2- and 3-defect configurations 
(not just those like in Fig.~\ref{fig1}(a)) are
included. The results of the two methods, therefore, should not be expected to
be identical. 

\section{Results}
\subsection{$B=0$ case }

\begin{figure}[t]
\includegraphics[width=0.9\columnwidth]{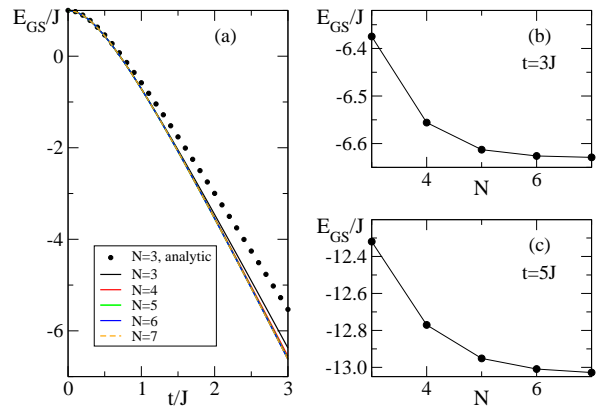}
\caption{(Color online) (a) GS energy vs. $t/J$, as predicted by the
  analytic solution (circles) and numerically in the variational
  spaces with $N=3,...,7$; (b) and (c) show the dependence on $N$ of
  the GS energy for $t/J=3$, respectively 5. 
\label{fig5}}
\end{figure}

We begin by looking at the
ground-state (GS)  energy as a function of $t/J$, using both the
analytical and the numerical methods. Results are shown in Fig.~\ref{fig5}(a) for $t/J\le 3$. As expected, in the limit
$t/J\rightarrow 0$, all curves converge towards $E_{\rm GS}=J=4\bj$, which
is the cost of placing the hole in the lattice (breaking four AFM
exchange bonds). As $t$ increases and the hole acquires a finite mass,
its energy is lowered. For small $t/J$, all values are in excellent
agreement, but for larger $t/J$ they begin to fan out. At $t/J=3$,
convergence has already been achieved for $N=7$, as shown in
Fig.~\ref{fig5}(b). For  $t/J=5$, $N=7$ is not yet fully
converged, although further  corrections are not expected to be
large. Note that we do not show results up to 
$t/J=5$ in panel (a) because the corresponding increase of the energy
range to be displayed makes it even more difficult to distinguish the numerical
curves from one another.

\begin{figure}[t]
\includegraphics[width=0.9\columnwidth]{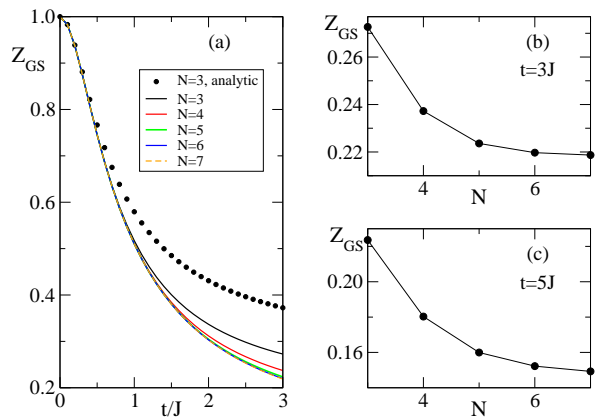}
\caption{(Color online) Same as in Fig.~\ref{fig5}, but for the
 GS quasiparticle weight instead of GS energy.
\label{fig6}}
\end{figure}

The analytical calculation indeed gives a reasonably accurate value up to
$t/J\sim 1\ldots 2$. The fact that the biggest variation is between it and
the $N=3$ numerical calculation, means that the 2- and 3-defect
configurations ignored in the analytical approximation play a
significant role in renormalizing the overall energy. This is not
surprising, since these are some of the least expensive
configurations. However, qualitatively and even quantitatively it is
clear that this 
analytical approximation is quite good up to fairly large $t/J$ values.

The quasiparticle weight 
\begin{equation}
Z({\mb k}) = | \langle
\phi_{\mb k}| h^\dagger_{\mb k}|{\rm GS}\rangle|^2\,,
\end{equation}
defined as the overlap between
the lowest eigenstate at a given momentum, 
\begin{equation}
{\cal H} |\phi_{\mb k}\rangle = E({\mb k}) |\phi_{\mb k}\rangle\,,
\end{equation}
 and the non-interacting
state $h^\dagger_{\mb k}|{\rm GS}\rangle$,  is shown at
${\mb k}=0$ in Fig.~\ref{fig6}. It remains quite considerable even for
large $t/J$, showing that a significant part of the wavefunction
consists of the hole alone, with no strings of defects. The effect of
increasing the variational space is more visible here, as expected due
to normalization: as the wavefunction acquires extra components in a
larger variational space, the weight of the hole-only part decreases,
even if overall the energy of the state is not much changed.

\begin{figure}[t]
\includegraphics[width=0.9\columnwidth]{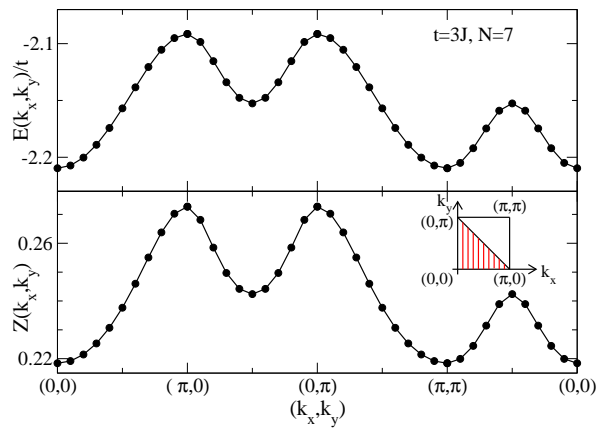}
\caption{(Color online) Energy $E({\mb k})/t$ (upper panel)
    and quasiparticle  
  weight $Z({\mb k})$ (lower panel) 
along the high-symmetry directions of the square lattice Brillouin zone,
  for $t/J=3$ and $N=7$. The inset in the bottom panel shows the upper
  right quadrant of the full Brillouin zone; the AFM Brillouin zone is
  shaded. Lines are guides to the eye.
\label{fig7}}
\end{figure}

The hole band dispersion $E({\mb k})$ and quasiparticle weight $Z({\mb
  k})$ are shown in Fig.~\ref{fig7}, along high-symmetry cuts in the
Brillouin zone of the original square lattice, for $t/J=3, N=7$. The
band folding due to the AFM order is clearly apparent.  These results
are very similar to those found for comparable parameters in the 2D
Edwards model (see Fig.~10 of Ref.~\onlinecite{BF10}): the minimum is
at ${\mb k}=(0,0)$, and ${\mb k}=({\pi\over2}, {\pi\over2})$ is a saddle
point (unlike in cuprates, where it is the actual minimum). The
quasiparticle weight is fairly constant but with local variations that
mimic the shape of the dispersion. This similarity is not surprising;
even though the Edwards model allows more states ({e.g.} with more
defects at a site), those are higher in energy and do not
contribute much to the GS. The mechanism for generating an effective
mass is identical in both models, with only small quantitative
differences due to the energy assigned to the strings of defects in
the two models.

\begin{figure}[t]
\includegraphics[width=0.9\columnwidth]{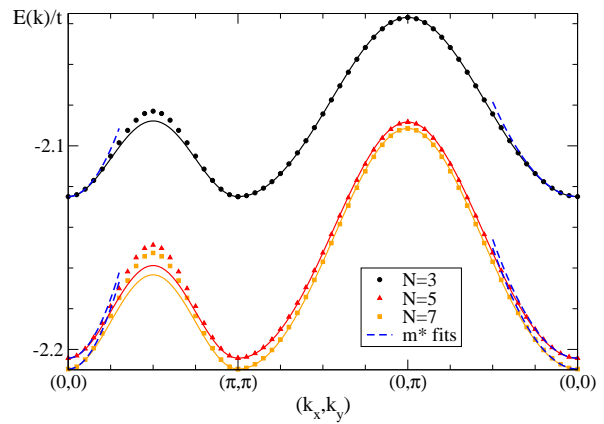}
\caption{(Color online) Dressed hole dispersion $E(\mb{k})/t$ along
  high-symmetry cuts in the Brillouin zone, for $t/J=3$ and
  $N=3,5,7$. Symbols are numerical data (for $N=7$, same as in
  Fig.~\ref{fig7}) and lines are results to a fit as in
  Eq.~(\ref{e1}), with parameters extracted for best fit on the
  $(0,0)-(\pi,0)$ cut. Dashed lines are fits to extract the effective mass.
\label{fig8}}
\end{figure}

One difference between the two models is illustrated in
Fig.~\ref{fig8}, where we show the dispersion for $N=3,5,7$ and $t/J=3$
along fewer cuts. The symbols are data from the numerical simulations,
and the lines are fits to an effective dispersion 
%%%%%%%%%%%%%%%%%%%%%%%%%%%%%% EQUATION %%%%%%%%%%%%%%%%%%%%%%%%%%%%%%
\begin{equation}
\EqLabel{e1}
E^*({\mb k}) = E_0 - 4 t_2^* \cos k_x  \cos k_y  -
2t_3^* \left[\cos(2k_x) + \cos(2k_y) \right],
\end{equation}
%%%%%%%%%%%%%%%%%%%%%%%%%%%%%%%%%%%%%%%%%%%%%%%%%%%%%%%%%%%%%%%%%%%%%%
along the $(0,0)\rightarrow (0,\pi)$ line. Using the same parameters
along the $(0,0)\rightarrow (\pi,\pi)$ line is clearly not a good
fit. In contrast, for the Edwards model such fits worked  well in
the entire Brillouin zone. This suggests that the retardation effects
may be somewhat stronger in this case: even though we know that only
$t_2(\omega)$ and $t_3(\omega)$ effective hoppings are generated, these
may not be well approximated by constant values $t_2^*, t_3^*$ if the
$\omega$ dependence is considerable within the bandwidth of the
dressed hole. 

This is less of an issue at smaller $t/J$, where the bandwidth is
significantly narrower.  In fact, even for $t/J=3$, the results for
$N=3$ shown in Fig.~\ref{fig8} clearly have a smaller bandwidth than
for $N=5,7$. This is not surprising, since longer loops with
additional contributions to effective hoppings are included in the
latter cases. Even though the decrease in bandwidth is not that large,
it is clear that the fit is better for the $N=3$ case. This may
explain why the fit worked well for the 2D Edwards model,\cite{BF10}
where the solution was restricted to $N=3$ and the bandwidths were
narrower also due to more costly defects (equivalent to larger $J$)
than in Fig.~\ref{fig8}.

\begin{figure}[t]
\includegraphics[width=0.9\columnwidth]{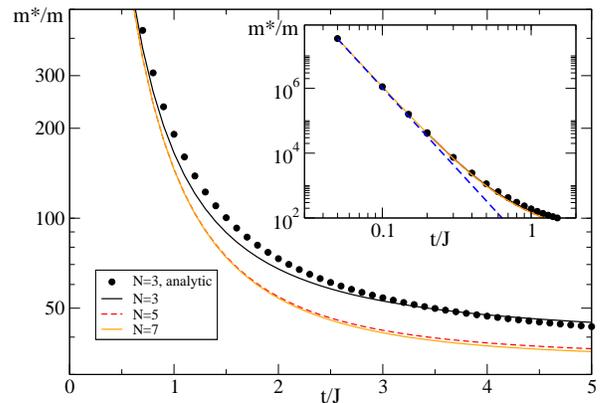}
\caption{(Color online) Effective mass vs. $t/J$. The inset shows the
  same data, plus the prediction of the weak-coupling perturbation
  theory (dashed line).
\label{fig9}}
\end{figure}

Since we cannot extract meaningful $t_2^*, t_3^*$ values, we instead
calculate the effective mass 
\begin{equation}
\left[\frac{1}{m^*}\right]_{x,y} = \left. \frac{\partial^2 E({\bf
    k})}{\partial k_x \partial k_y} \right|_{|{\bf k}|\to 0}\,, 
\end{equation}
with fits as shown by the dashed lines in Fig.~\ref{fig8}. As
expected, the mass is found to be isotropic. The results are shown in
Fig.~\ref{fig9}, on a logarithmic scale. The effective mass $m^*$ is
in units of the band mass $m$, i.e. the bare particle mass if there
was no  AFM background.

We show results up to $t/J=5$, even though the values are not
fully converged for $t/J>3$. At these larger values,
longer loops than $N=7$ would need to be included. From results such as in
Fig.~\ref{fig5}(c), however, we do not expect those further
corrections to be very significant. Also, they would decrease $m^*$, as they
would open additional channels for effective hopping, so the values
shown in Fig.~\ref{fig9} can be taken as an upper bound for
$m^*$. 

In the limit $t/J \rightarrow 0$, the effective mass becomes very
large. As discussed, from perturbation theory here we expect $t_2 =
t^6/(21600 \bj^5) $. The corresponding $m^*/m = 10.55 (J/t)^5$ is shown
as a dashed line in the inset, in good agreement with the values
obtained from the analytical and numerical calculations (symbols and
full lines). The effective mass diverges as $t/J\to0$ since 
the hole is bound
to its original site in this case. As $t/J$ increases, the effective
mass decreases significantly, and it reaches $m^*/m \sim 40$ for
$t/J=3$. Since this decrease is mirrored by the analytical $N=3$
calculation, it must be due to the summation over repeated 3-defect
loops, like those shown in Fig.~\ref{fig2}. Longer loops further lower the
effective mass. The $N=5$ loops give a significant contribution for
$t/J>1$, validating the expectations based on perturbation theory. The
contribution of the $N=7$ loops is still quite small, for these
values. We expect $m^*$ to further decrease with
increasing $t/J \rightarrow \infty$. However, capturing that limit is
very difficult if not outright impossible with this variational
formulation, given the huge increase in possible configurations with increasing
 $N$.

The main result, thus far, is that a hole in an Ising AFM is
fairly mobile if $t$ and $J$ are comparable, even though there
are no spin fluctuations in this model.

\subsection{$B\neq 0$ case}

We now proceed to discuss the spectrum of the dressed hole in the
presence of a transverse magnetic field $B$. We will use the $N=3$
analytical Eq.~(\ref{n3}), and solve it numerically for various values
of $\phi = \pi p/n$. We show results for $t/J=3$, even though we know
that here the analytical calculation is not sufficient for full
convergence, simply because its $B=0$ bandwidth is sufficiently large to make
it easier to see the effect of a finite $B$. We note that the
numerical calculation in the variational spaces 
corresponding to various $N$ can be carried out as well, however each
previous unknown, corresponding to an allowed configuration of
defects, now becomes a $2n \times 2n$ matrix of unknowns,
corresponding to each of the $2n$ distinct sites in the magnetic
Brillouin zone shown in Fig.~\ref{fig3}. This huge increase in the size
of the linear system to be solved, especially for larger $n$ values
(smaller magnetic fields) makes the implementation of this scheme
cumbersome, and unnecessary since we do not expect any
qualitative changes.

\begin{figure*}[t]
\includegraphics[width=2\columnwidth]{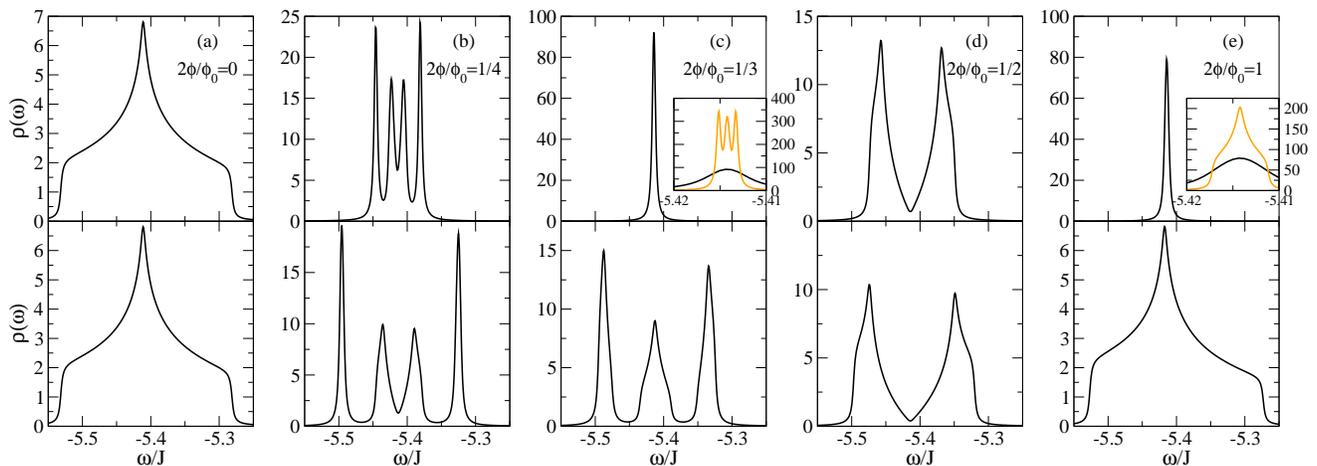}
\caption{(Color online) Local density of states $\rho(\omega)$
  vs. $\omega/J$ for various magnetic fluxes through the sublattice
  unit cell $2\Phi/\Phi_0=0, {1\over4}, {1\over 3}, {1\over 2}$ and $ 1$, for
  $t/J=3$. The upper panels show the full solution of Eq.~(\ref{n3}),
  while the lower panels show the solution if the Aharonov-Bohm
  interference is turned off by setting $\phi=0$ in
  Eqs.~(\ref{n4})-(\ref{n6}). The broadening is 
  $\eta/J=2.5\cdot 10^{-3}$, except for the orange (light) curves in
  the insets, for which $\eta/J=2.5\cdot 10^{-4}$.
\label{fig10}}
\end{figure*}

After solving the linear system in
Eq.~(\ref{n3}), we calculate the local density of states (LDOS)
%%%%%%%%%%%%%%%%%%%%%%%%%%%%%% EQUATION %%%%%%%%%%%%%%%%%%%%%%%%%%%%%%
\begin{equation}
\EqLabel{rho2}
\rho(\omega) = -{1\over \pi} \mbox{Im } \sum_{\mb k}^{}
G_{\alpha\alpha}({\mb k}, \omega),
\end{equation}
%%%%%%%%%%%%%%%%%%%%%%%%%%%%%%%%%%%%%%%%%%%%%%%%%%%%%%%%%%%%%%%%%%%%%%
where the sum is over the (highly folded) magnetic Brillouin zone
corresponding to the magnetic unit cell of Fig.~\ref{fig3}. This LDOS
is independent of the site $\alpha=0,...,2n-1$ used, hence the lack of
a site index. 

We first consider the effect of the additional dependence on $B$
due to the  Aharonov-Bohm interference, on the Hofstadter butterfly
expected when the magnetic flux through the unit cell is comparable to
$\Phi_0$. Results for various simple ratios are shown in
Fig.~\ref{fig10}. In all cases, the upper panel shows the full
results, whereas the lower panel shows the results when the Aharonov-Bohm 
interference is turned off, {i.e.} we use the $\phi=0$ values  
$\epsilon_0(\phi=0,\omega)$, $t_2(\phi=0,\omega)$ and
$t_3(\phi=0,\omega)$ in Eq.~(\ref{n3}).

For the $B=0$ case shown in (a), the results are of course identical. The
density of states resembles that of a 2D square lattice with NN
hopping. This is expected, since the dominant hopping term is $t_2$,
which plays the role of NN hopping for the sublattice on which the
dressed hole lives. The LDOS is slightly distorted due to (small)
contributions from the $t_3$ term and the retardation effects. These
small distortions and asymmetries are observed in all other panels.

For a finite flux through the sublattice unit cell $2\Phi/\Phi_0=1/n$,
the band splits into $n$ subbands, as expected in standard Hofstadter
butterfly phenomenology. The bands are not fully separated because we
used a broadening $\eta/J=2.5\cdot 10^{-3}$ which, while small,
is still comparable with some of these features' bandwidth. Despite
this, the various subbands are easily identifiable.

In Figs.~\ref{fig10}~(b) and~(d), the full result (top panel), while
quite similar in aspect to the lower panel, shows a somewhat narrower
bandwidth. This is not surprising, since the Aharonov-Bohm interference terms
like $\cos(3\phi/2)$ etc. are responsible for a decrease in the value
of the effective hoppings, and therefore of the overall
bandwidth. This also explains the apparent ``collapse'' of the
butterfly for $\phi = \pi/3$ and $\phi=\pi$, shown in Figs.~\ref{fig10}
(c) and (e), respectively. The very narrow peak seen in both cases
shows the expected subband structure, if a much smaller
$\eta/J=2.5\cdot 10^{-4}$ values is used, as done in the inset: the
former case shows the 3 subbands (these features are so narrow that
even this much smaller $\eta$ can only partially resolve them), while
the later case shows the one band. The extreme narrowing is due to the
Aharonov-Bohm interference, which at these values of the flux results in a very
small or vanishing $t_2(\phi, \omega)$, see Eq.~(\ref{n5}), and a
bandwidth set by $t_3(\phi, \omega)\ll t_2(0,\omega)$.

As already noted, this additional $B$-dependent modulation of the
Hofstadter structure, coming from the Aharonov-Bohm interference, is
also responsible for an increase in the periodicity of the
butterfly. The lower panel shows that, as expected in models with
constant hopping integrals, the butterfly is periodic if the flux
through the unit cell increases by a flux quantum (here, $2\Phi
\rightarrow 2\Phi + \Phi_0$, or $\phi \rightarrow \phi+\pi$). The full
result clearly does not have this periodicity. As discussed
previously, based on Eqs.~(\ref{n4})--(\ref{n6}) we expect the
periodicity to be $\phi \rightarrow \phi+2\pi$ for the full
Aharonov-Bohm case; our simulations confirm this (not shown).

We can imagine that this $2\pi$ periodicity 
survives if one allows longer
loops in the calculation, but some care is needed. For example,
loops like those shown in Fig.~\ref{fig1}(b) would bring in factors of
$\cos(7\phi/2)$ in their contribution to $t_2$, and
various other Aharonov-Bohm phases will be associated with longer loops. We
calculated several of these and all are consistent with the $2\pi$
periodicity; so we believe this $2\pi$ period is correct to all
orders, but do not have a 
proof.

Finally, we discuss the role of the Aharonov-Bohm interference at very small
magnetic fields. In this case, one expects the spectrum to separate in
a sequence of Landau levels (LLs) separated by the cyclotron
frequency. The appearance of LLs in the spectrum leads to
 quantum oscillations in various transport measurements,
such as de Haas-van Alphen oscillations.

\begin{figure}[t]
\includegraphics[width=\columnwidth]{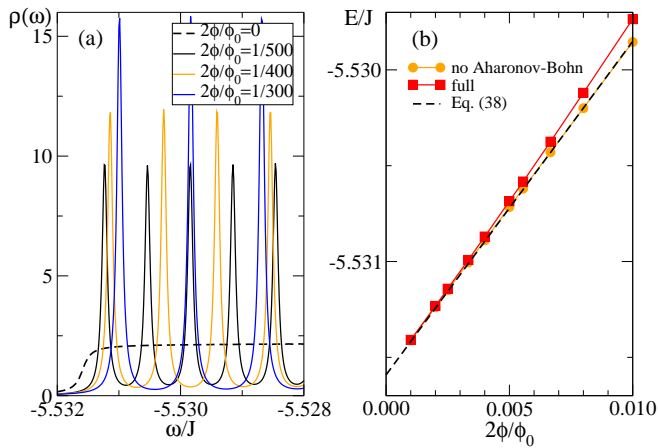}
\caption{(Color online) (a) LDOS for small magnetic fields (full
  lines), showing several LLs at the bottom of the spectrum. The
  dashed lines shows the $B=0$ LDOS; (b) the location of the lowest LL peak,
 in the full calculation (squares) and with the Aharonov-Bohm interference turned
  off (circles). The dashed line is the analytical prediction of
  Eq.~(\ref{27}). These results are for  $t/J=3$ and $\eta/J =
  10^{-5}$. 
\label{fig11}}
\end{figure}

The LDOS at the lower edge, with (full lines) and without (dashed
line) very small magnetic fields applied, is shown in
Fig.~\ref{fig11}(a). The $B=0$ LDOS shows the jump at the band edge, here
smoothed out by a finite $\eta$ value, and the roughly constant 2D
density of states above it. The slight monotonic increase is due to
the use of a tight-binding dispersion, as opposed to an approximate
parabolic one.\cite{BC10} When the small field is turned on, we see
that this LDOS splits into equally spaced narrow peaks, marking the
LLs. The cyclotron frequency, defined by the distance between
consecutive LLs, increases roughly linearly with $B$, and so does the
weight in each LLs, as expected because of its larger degeneracy.

In Fig.~\ref{fig11}(b), we plot the energy of the lowest LL 
against the flux through the unit cell. Squares show the full result,
while circles are the result when the Aharonov-Bohm oscillations are turned
off. The dashed line shows the value of the lowest LL for a particle of
constant mass $m^*$, 
%%%%%%%%%%%%%%%%%%%%%%%%%%%%%% EQUATION %%%%%%%%%%%%%%%%%%%%%%%%%%%%%%
\begin{equation}
\EqLabel{27}
\varepsilon = [E_{\rm GS} + \hbar \omega_c/ 2]/J\,,
\end{equation}
%%%%%%%%%%%%%%%%%%%%%%%%%%%%%%%%%%%%%%%%%%%%%%%%%%%%%%%%%%%%%%%%%%%%%%
where the cyclotron frequency is $\omega_c= eB/m^*$, and $E_{\rm GS}$ and
$m^*$ are the $B=0$ values of the dressed hole ground-state energy and
effective mass, respectively. This prediction is in excellent
agreement with the results where the Aharonov-Bohm interference 
is turned off. In 
contrast, the full results show additional quadratic
$B$-dependence. This is not so surprising, since for small magnetic
fields, the effective hoppings depend quadratically on $B$ through the
Aharonov-Bohm interference terms, and  one would expect that dependence
to be mirrored here through the effective mass.

We do not perform a more quantitative analysis because, as already
mentioned, the $N=3$ analytical case is not fully converged at this
value of $t/J=3$; for example, the effective mass is $m^*/m \approx
54$, whereas the converged result is $m^*/m \approx
40$ -- this is quite a sizable difference. Going to a smaller $t/J$ is
not easy either, because the bandwidth narrows (or, equally, the effective
mass increases) and it becomes more and more difficult to separate
various features.\\

\section{Summary}

To conclude, we have investigated the motion of a hole in a 2D square
Ising AFM. We showed that summation of the contribution of all Trugman
loops up to a given length $N$ can be carried out numerically, and
convergence is reached for $N=7$ if $t/J\lesssim
  3$. Qualitatively correct 
and quantitatively quite reasonably accurate results are obtained from
a simple analytical approximation for $N=3$. We find that the
effective mass of the hole can be fairly low, of around 30\ldots40~$m$, if
$t/J\sim$~3\ldots5.

If a magnetic field is turned on, Aharonov-Bohm interference between
clockwise and counterclockwise Trugman loops leads to dependence of
the effective hoppings on the magnetic field over and above the usual
Peierls phases. Their most spectacular manifestation is in the
``collapse'' of the Hofstadter butterfly structure at fields where the
Aharonov-Bohm interference is destructive. Of course, creating large enough
magnetic fields to see the Hofstadter butterfly band structure in a
crystal is still a challenge (moreover, one would need a large $J$ in
order to prevent a transition to ferromagnet order, at such large
fields). Nevertheless, this is an interesting effect that might be
mimicked in some other type of systems.

For small magnetic fields, where the magnetic length is large as
compared to the unit cell, Landau levels form. Here, we find that the
effect of the Aharonov-Bohm phases is to bring additional dependence on $B$
through $m^*(B)$, in the cyclotron frequency. This is a small effect,
but one that might be easier to see through various quantum
oscillations-type measurements.

\begin{acknowledgments}
We thank A. Alvermann, A. Aharony, D. M. Edwards, O. Entin-Wohlman and
G. Sawatzky for useful suggestions and
discussions.
This work was supported by NSERC and CIFAR (MB) and DFG SFB 652
(HF). 
\end{acknowledgments}

%\bibliographystyle{apsrev4-1}
%\bibliography{ref} 

\begin{thebibliography}{25}%
\makeatletter
\providecommand \@ifxundefined [1]{%
 \@ifx{#1\undefined}
}%
\providecommand \@ifnum [1]{%
 \ifnum #1\expandafter \@firstoftwo
 \else \expandafter \@secondoftwo
 \fi
}%
\providecommand \@ifx [1]{%
 \ifx #1\expandafter \@firstoftwo
 \else \expandafter \@secondoftwo
 \fi
}%
\providecommand \natexlab [1]{#1}%
\providecommand \enquote  [1]{``#1''}%
\providecommand \bibnamefont  [1]{#1}%
\providecommand \bibfnamefont [1]{#1}%
\providecommand \citenamefont [1]{#1}%
\providecommand \href@noop [0]{\@secondoftwo}%
\providecommand \href [0]{\begingroup \@sanitize@url \@href}%
\providecommand \@href[1]{\@@startlink{#1}\@@href}%
\providecommand \@@href[1]{\endgroup#1\@@endlink}%
\providecommand \@sanitize@url [0]{\catcode `\\12\catcode `\$12\catcode
  `\&12\catcode `\#12\catcode `\^12\catcode `\_12\catcode `\%12\relax}%
\providecommand \@@startlink[1]{}%
\providecommand \@@endlink[0]{}%
\providecommand \url  [0]{\begingroup\@sanitize@url \@url }%
\providecommand \@url [1]{\endgroup\@href {#1}{\urlprefix }}%
\providecommand \urlprefix  [0]{URL }%
\providecommand \Eprint [0]{\href }%
\providecommand \doibase [0]{http://dx.doi.org/}%
\providecommand \selectlanguage [0]{\@gobble}%
\providecommand \bibinfo  [0]{\@secondoftwo}%
\providecommand \bibfield  [0]{\@secondoftwo}%
\providecommand \translation [1]{[#1]}%
\providecommand \BibitemOpen [0]{}%
\providecommand \bibitemStop [0]{}%
\providecommand \bibitemNoStop [0]{.\EOS\space}%
\providecommand \EOS [0]{\spacefactor3000\relax}%
\providecommand \BibitemShut  [1]{\csname bibitem#1\endcsname}%
\let\auto@bib@innerbib\@empty
%</preamble>
\bibitem [{\citenamefont {Bednorz}\ and\ \citenamefont
  {M\"uller}(1986)}]{BM86}%
  \BibitemOpen
  \bibfield  {author} {\bibinfo {author} {\bibfnamefont {I.~G.}\ \bibnamefont
  {Bednorz}}\ and\ \bibinfo {author} {\bibfnamefont {K.~A.}\ \bibnamefont
  {M\"uller}},\ }\href@noop {} {\bibfield  {journal} {\bibinfo  {journal} {Z.
  Phys. B}\ }\textbf {\bibinfo {volume} {64}},\ \bibinfo {pages} {189}
  (\bibinfo {year} {1986})}\BibitemShut {NoStop}%
\bibitem [{\citenamefont {Berciu}(2009)}]{Be09}%
  \BibitemOpen
  \bibfield  {author} {\bibinfo {author} {\bibfnamefont {M.}~\bibnamefont
  {Berciu}},\ }\href@noop {} {\bibfield  {journal} {\bibinfo  {journal}
  {Physics}\ }\textbf {\bibinfo {volume} {2}},\ \bibinfo {pages} {55} (\bibinfo
  {year} {2009})}\BibitemShut {NoStop}%
\bibitem [{\citenamefont {Hubbard}(1963)}]{Hu63}%
  \BibitemOpen
  \bibfield  {author} {\bibinfo {author} {\bibfnamefont {J.}~\bibnamefont
  {Hubbard}},\ }\href@noop {} {\bibfield  {journal} {\bibinfo  {journal} {Proc.
  Roy. Soc. London, Ser. A}\ }\textbf {\bibinfo {volume} {276}},\ \bibinfo
  {pages} {238} (\bibinfo {year} {1963})}\BibitemShut {NoStop}%
\bibitem [{\citenamefont {Kanamori}(1963)}]{Ka63}%
  \BibitemOpen
  \bibfield  {author} {\bibinfo {author} {\bibfnamefont {J.}~\bibnamefont
  {Kanamori}},\ }\href@noop {} {\bibfield  {journal} {\bibinfo  {journal}
  {Prog. Theor. Phys.}\ }\textbf {\bibinfo {volume} {30}},\ \bibinfo {pages}
  {275} (\bibinfo {year} {1963})}\BibitemShut {NoStop}%
\bibitem [{\citenamefont {Gros}\ \emph {et~al.}(1987)\citenamefont {Gros},
  \citenamefont {Joynt},\ and\ \citenamefont {Rice}}]{GJR87}%
  \BibitemOpen
  \bibfield  {author} {\bibinfo {author} {\bibfnamefont {C.}~\bibnamefont
  {Gros}}, \bibinfo {author} {\bibfnamefont {M.~R.}\ \bibnamefont {Joynt}}, \
  and\ \bibinfo {author} {\bibfnamefont {T.~M.}\ \bibnamefont {Rice}},\
  }\href@noop {} {\bibfield  {journal} {\bibinfo  {journal} {Phys. Rev. B}\
  }\textbf {\bibinfo {volume} {36}},\ \bibinfo {pages} {3583} (\bibinfo {year}
  {1987})}\BibitemShut {NoStop}%
\bibitem [{\citenamefont {Trugman}(1988)}]{Tr88}%
  \BibitemOpen
  \bibfield  {author} {\bibinfo {author} {\bibfnamefont {S.~A.}\ \bibnamefont
  {Trugman}},\ }\href@noop {} {\bibfield  {journal} {\bibinfo  {journal} {Phys.
  Rev. B}\ }\textbf {\bibinfo {volume} {37}},\ \bibinfo {pages} {1597}
  (\bibinfo {year} {1988})}\BibitemShut {NoStop}%
\bibitem [{\citenamefont {Kane}\ \emph {et~al.}(1989)\citenamefont {Kane},
  \citenamefont {Lee},\ and\ \citenamefont {Read}}]{KLR89}%
  \BibitemOpen
  \bibfield  {author} {\bibinfo {author} {\bibfnamefont {C.~L.}\ \bibnamefont
  {Kane}}, \bibinfo {author} {\bibfnamefont {P.~A.}\ \bibnamefont {Lee}}, \
  and\ \bibinfo {author} {\bibfnamefont {N.}~\bibnamefont {Read}},\ }\href@noop
  {} {\bibfield  {journal} {\bibinfo  {journal} {Phys. Rev. B}\ }\textbf
  {\bibinfo {volume} {39}},\ \bibinfo {pages} {6880} (\bibinfo {year}
  {1989})}\BibitemShut {NoStop}%
\bibitem [{\citenamefont {Martinez}\ and\ \citenamefont
  {Horsch}(1991)}]{MH91a}%
  \BibitemOpen
  \bibfield  {author} {\bibinfo {author} {\bibfnamefont {G.}~\bibnamefont
  {Martinez}}\ and\ \bibinfo {author} {\bibfnamefont {P.}~\bibnamefont
  {Horsch}},\ }\href@noop {} {\bibfield  {journal} {\bibinfo  {journal} {Phys.
  Rev. B}\ }\textbf {\bibinfo {volume} {44}},\ \bibinfo {pages} {317} (\bibinfo
  {year} {1991})}\BibitemShut {NoStop}%
\bibitem [{\citenamefont {Dagotto}\ \emph {et~al.}(1990)\citenamefont
  {Dagotto}, \citenamefont {Joynt}, \citenamefont {Moreo}, \citenamefont
  {Bacci},\ and\ \citenamefont {Gagliano}}]{Daea90}%
  \BibitemOpen
  \bibfield  {author} {\bibinfo {author} {\bibfnamefont {E.}~\bibnamefont
  {Dagotto}}, \bibinfo {author} {\bibfnamefont {R.}~\bibnamefont {Joynt}},
  \bibinfo {author} {\bibfnamefont {A.}~\bibnamefont {Moreo}}, \bibinfo
  {author} {\bibfnamefont {S.}~\bibnamefont {Bacci}}, \ and\ \bibinfo {author}
  {\bibfnamefont {E.}~\bibnamefont {Gagliano}},\ }\href@noop {} {\bibfield
  {journal} {\bibinfo  {journal} {Phys. Rev. B}\ }\textbf {\bibinfo {volume}
  {41}},\ \bibinfo {pages} {9049} (\bibinfo {year} {1990})}\BibitemShut
  {NoStop}%
\bibitem [{\citenamefont {Fehske}\ \emph {et~al.}(1991)\citenamefont {Fehske},
  \citenamefont {Waas}, \citenamefont {R\"oder},\ and\ \citenamefont
  {B\"uttner}}]{FWRB91}%
  \BibitemOpen
  \bibfield  {author} {\bibinfo {author} {\bibfnamefont {H.}~\bibnamefont
  {Fehske}}, \bibinfo {author} {\bibfnamefont {V.}~\bibnamefont {Waas}},
  \bibinfo {author} {\bibfnamefont {H.}~\bibnamefont {R\"oder}}, \ and\
  \bibinfo {author} {\bibfnamefont {H.}~\bibnamefont {B\"uttner}},\ }\href@noop
  {} {\bibfield  {journal} {\bibinfo  {journal} {Phys. Rev. B}\ }\textbf
  {\bibinfo {volume} {44}},\ \bibinfo {pages} {8473} (\bibinfo {year}
  {1991})}\BibitemShut {NoStop}%
\bibitem [{\citenamefont {Dagotto}(1994)}]{Da94}%
  \BibitemOpen
  \bibfield  {author} {\bibinfo {author} {\bibfnamefont {E.}~\bibnamefont
  {Dagotto}},\ }\href@noop {} {\bibfield  {journal} {\bibinfo  {journal} {Rev.
  Mod. Phys.}\ }\textbf {\bibinfo {volume} {66}},\ \bibinfo {pages} {763}
  (\bibinfo {year} {1994})}\BibitemShut {NoStop}%
\bibitem [{\citenamefont {Lee}\ \emph {et~al.}(2006)\citenamefont {Lee},
  \citenamefont {Nagaosa},\ and\ \citenamefont {Wen}}]{LNW06}%
  \BibitemOpen
  \bibfield  {author} {\bibinfo {author} {\bibfnamefont {P.~A.}\ \bibnamefont
  {Lee}}, \bibinfo {author} {\bibfnamefont {N.}~\bibnamefont {Nagaosa}}, \ and\
  \bibinfo {author} {\bibfnamefont {X.-G.}\ \bibnamefont {Wen}},\ }\href@noop
  {} {\bibfield  {journal} {\bibinfo  {journal} {Rev. Mod. Phys.}\ }\textbf
  {\bibinfo {volume} {78}},\ \bibinfo {pages} {17} (\bibinfo {year}
  {2006})}\BibitemShut {NoStop}%
\bibitem [{\citenamefont {Lee}(2008)}]{Le08}%
  \BibitemOpen
  \bibfield  {author} {\bibinfo {author} {\bibfnamefont {P.~A.}\ \bibnamefont
  {Lee}},\ }\href@noop {} {\bibfield  {journal} {\bibinfo  {journal} {Rep.
  Prog. Phys.}\ }\textbf {\bibinfo {volume} {71}},\ \bibinfo {pages} {012501}
  (\bibinfo {year} {2008})}\BibitemShut {NoStop}%
\bibitem [{\citenamefont {Shraiman}\ and\ \citenamefont {Sigga}(1988)}]{SS88}%
  \BibitemOpen
  \bibfield  {author} {\bibinfo {author} {\bibfnamefont {B.~I.}\ \bibnamefont
  {Shraiman}}\ and\ \bibinfo {author} {\bibfnamefont {E.~D.}\ \bibnamefont
  {Sigga}},\ }\href@noop {} {\bibfield  {journal} {\bibinfo  {journal} {Phys.
  Rev. Lett.}\ }\textbf {\bibinfo {volume} {60}},\ \bibinfo {pages} {740}
  (\bibinfo {year} {1988})}\BibitemShut {NoStop}%
\bibitem [{\citenamefont {Eder}\ \emph {et~al.}(1990)\citenamefont {Eder},
  \citenamefont {Becker},\ and\ \citenamefont {Stephan}}]{EBS90}%
  \BibitemOpen
  \bibfield  {author} {\bibinfo {author} {\bibfnamefont {R.}~\bibnamefont
  {Eder}}, \bibinfo {author} {\bibfnamefont {K.~W.}\ \bibnamefont {Becker}}, \
  and\ \bibinfo {author} {\bibfnamefont {W.~H.}\ \bibnamefont {Stephan}},\
  }\href@noop {} {\bibfield  {journal} {\bibinfo  {journal} {Z. Phys. B}\
  }\textbf {\bibinfo {volume} {81}},\ \bibinfo {pages} {33} (\bibinfo {year}
  {1990})}\BibitemShut {NoStop}%
\bibitem [{\citenamefont {Brinkman}\ and\ \citenamefont {Rice}(1970)}]{BR70}%
  \BibitemOpen
  \bibfield  {author} {\bibinfo {author} {\bibfnamefont {W.~F.}\ \bibnamefont
  {Brinkman}}\ and\ \bibinfo {author} {\bibfnamefont {T.~M.}\ \bibnamefont
  {Rice}},\ }\href@noop {} {\bibfield  {journal} {\bibinfo  {journal} {Phys.
  Rev. B}\ }\textbf {\bibinfo {volume} {2}},\ \bibinfo {pages} {4302} (\bibinfo
  {year} {1970})}\BibitemShut {NoStop}%
\bibitem [{\citenamefont {Edwards}(2006)}]{Ed06}%
  \BibitemOpen
  \bibfield  {author} {\bibinfo {author} {\bibfnamefont {D.~M.}\ \bibnamefont
  {Edwards}},\ }\href@noop {} {\bibfield  {journal} {\bibinfo  {journal}
  {Physica B}\ }\textbf {\bibinfo {volume} {378-380}},\ \bibinfo {pages} {133}
  (\bibinfo {year} {2006})}\BibitemShut {NoStop}%
\bibitem [{\citenamefont {Alvermann}\ \emph {et~al.}(2007)\citenamefont
  {Alvermann}, \citenamefont {Edwards},\ and\ \citenamefont {Fehske}}]{AEF07}%
  \BibitemOpen
  \bibfield  {author} {\bibinfo {author} {\bibfnamefont {A.}~\bibnamefont
  {Alvermann}}, \bibinfo {author} {\bibfnamefont {D.~M.}\ \bibnamefont
  {Edwards}}, \ and\ \bibinfo {author} {\bibfnamefont {H.}~\bibnamefont
  {Fehske}},\ }\href@noop {} {\bibfield  {journal} {\bibinfo  {journal} {Phys.
  Rev. Lett.}\ }\textbf {\bibinfo {volume} {98}},\ \bibinfo {pages} {056602}
  (\bibinfo {year} {2007})}\BibitemShut {NoStop}%
\bibitem [{\citenamefont {Edwards}\ \emph {et~al.}(2010)\citenamefont
  {Edwards}, \citenamefont {Ejima}, \citenamefont {Alvermann},\ and\
  \citenamefont {Fehske}}]{EEAF10}%
  \BibitemOpen
  \bibfield  {author} {\bibinfo {author} {\bibfnamefont {D.~M.}\ \bibnamefont
  {Edwards}}, \bibinfo {author} {\bibfnamefont {S.}~\bibnamefont {Ejima}},
  \bibinfo {author} {\bibfnamefont {A.}~\bibnamefont {Alvermann}}, \ and\
  \bibinfo {author} {\bibfnamefont {H.}~\bibnamefont {Fehske}},\ }\href@noop {}
  {\bibfield  {journal} {\bibinfo  {journal} {J. Phys. Condens. Matter}\
  }\textbf {\bibinfo {volume} {22}},\ \bibinfo {pages} {435601} (\bibinfo
  {year} {2010})}\BibitemShut {NoStop}%
\bibitem [{\citenamefont {Alvermann}\ \emph {et~al.}(2010)\citenamefont
  {Alvermann}, \citenamefont {Edwards},\ and\ \citenamefont {Fehske}}]{AEF10}%
  \BibitemOpen
  \bibfield  {author} {\bibinfo {author} {\bibfnamefont {A.}~\bibnamefont
  {Alvermann}}, \bibinfo {author} {\bibfnamefont {D.~M.}\ \bibnamefont
  {Edwards}}, \ and\ \bibinfo {author} {\bibfnamefont {H.}~\bibnamefont
  {Fehske}},\ }\href@noop {} {\bibfield  {journal} {\bibinfo  {journal} {J.
  Phys. Conf. Ser.}\ }\textbf {\bibinfo {volume} {220}},\ \bibinfo {pages}
  {012023} (\bibinfo {year} {2010})}\BibitemShut {NoStop}%
\bibitem [{\citenamefont {Berciu}\ and\ \citenamefont {Fehske}(2010)}]{BF10}%
  \BibitemOpen
  \bibfield  {author} {\bibinfo {author} {\bibfnamefont {M.}~\bibnamefont
  {Berciu}}\ and\ \bibinfo {author} {\bibfnamefont {H.}~\bibnamefont
  {Fehske}},\ }\href@noop {} {\bibfield  {journal} {\bibinfo  {journal} {Phys.
  Rev. B}\ }\textbf {\bibinfo {volume} {82}},\ \bibinfo {pages} {085116}
  (\bibinfo {year} {2010})}\BibitemShut {NoStop}%
\bibitem [{\citenamefont {Hofstadter}(1976)}]{Ho76}%
  \BibitemOpen
  \bibfield  {author} {\bibinfo {author} {\bibfnamefont {D.~R.}\ \bibnamefont
  {Hofstadter}},\ }\href@noop {} {\bibfield  {journal} {\bibinfo  {journal}
  {Phys. Rev. B}\ }\textbf {\bibinfo {volume} {14}},\ \bibinfo {pages} {2239}
  (\bibinfo {year} {1976})}\BibitemShut {NoStop}%
\bibitem [{\citenamefont {Schenk}\ and\ \citenamefont
  {G\"artner}(2004)}]{SG04}%
  \BibitemOpen
  \bibfield  {author} {\bibinfo {author} {\bibfnamefont {O.}~\bibnamefont
  {Schenk}}\ and\ \bibinfo {author} {\bibfnamefont {K.}~\bibnamefont
  {G\"artner}},\ }\href@noop {} {\bibfield  {journal} {\bibinfo  {journal}
  {Journal Future Generation Computer Systems}\ }\textbf {\bibinfo {volume}
  {20}},\ \bibinfo {pages} {475} (\bibinfo {year} {2004})}\BibitemShut
  {NoStop}%
\bibitem [{\citenamefont {Schenk}\ and\ \citenamefont
  {G\"artner}(2006)}]{SG06}%
  \BibitemOpen
  \bibfield  {author} {\bibinfo {author} {\bibfnamefont {O.}~\bibnamefont
  {Schenk}}\ and\ \bibinfo {author} {\bibfnamefont {K.}~\bibnamefont
  {G\"artner}},\ }\href@noop {} {\bibfield  {journal} {\bibinfo  {journal}
  {Elec. Trans. Numer. Anal.}\ }\textbf {\bibinfo {volume} {23}},\ \bibinfo
  {pages} {158} (\bibinfo {year} {2006})}\BibitemShut {NoStop}%
\bibitem [{\citenamefont {Berciu}\ and\ \citenamefont {Cook}(2010)}]{BC10}%
  \BibitemOpen
  \bibfield  {author} {\bibinfo {author} {\bibfnamefont {M.}~\bibnamefont
  {Berciu}}\ and\ \bibinfo {author} {\bibfnamefont {A.~M.}\ \bibnamefont
  {Cook}},\ }\href@noop {} {\bibfield  {journal} {\bibinfo  {journal}
  {Europhys. Lett.}\ }\textbf {\bibinfo {volume} {92}},\ \bibinfo {pages}
  {40003} (\bibinfo {year} {2010})}\BibitemShut {NoStop}%
\end{thebibliography}
%merlin.mbs apsrev4-1.bst 2010-07-25 4.21a (PWD, AO, DPC) hacked
%Control: key (0)
%Control: author (72) initials jnrlst
%Control: editor formatted (1) identically to author
%Control: production of article title (-1) disabled
%Control: page (0) single
%Control: year (1) truncated
%Control: production of eprint (0) enabled
%

\end{document}